%% file: IEEE_paper_main.tex
\documentclass[%
	final,
	conference, 
    comsoc,
	letterpaper,
	oneside,
	twocolumn,
	nofonttune,%
]{IEEEtran}%
\input{./organization/preamble.tex}%
\input{./organization/makros.tex}%
\input{./organization/settings.tex}%
\makeatletter
\let\blx@rerun@biber\relax
\makeatother

\makeatletter
    \let\thanks\@IEEESAVECMDthanks%
\makeatother

\usepackage{listings}
\usepackage[usenames,dvipsnames]{xcolor}
\usepackage{tikz} \usetikzlibrary{calc, arrows.meta, intersections, patterns, positioning, shapes.misc, fadings, through,decorations.pathreplacing}

\definecolor{ColorOne}{named}{MidnightBlue}
\definecolor{ColorTwo}{named}{Dandelion}
\definecolor{ColorThree}{named}{Plum}

\usepackage{algorithm}
\usepackage{algpseudocode}
\usepackage{tabularx}
\usepackage{newtxmath}

\usepackage{booktabs}
\pgfplotsset{
  grid style = {
   line width = 0.1pt
  }
}
				\newcommand{\disablewr}[1]{#1}%
				\newcommand{\newcommanddisw}[3]{\newcommand{#1}[1]{\disablewr{\textcolor{#2}{#3}}}}%
\renewcommand{\disablewr}[1]{}%
\definecolor{todocol}{named}{red}
\newcommanddisw{\todo}{todocol}{ToDo: #1}%
\definecolor{migucol}{named}{purple}%
\newcommanddisw{\migucom}{migucol}{{@}comment: #1}%
\newcommanddisw{\miguhigh}{migucol}{#1}%
\usepackage{siunitx}
\usepackage{xspace}

\begin{document}%
%
\title{%
Benchmarking OpenWiFiSync on ESP32: Towards Cost-Effective Wireless Time Synchronization
\thanks{This research was supported by the German Federal Ministry of Research, Technology and Space (BMFTR) within the Open6GHub and the 6G-Health project under grant numbers 16KISK003K and 16KISK214. This is a preprint of a work accepted but not yet published at the 30th IEEE International Conference on Emerging Technologies and Factory Automation (ETFA). Please cite as: M.~Gundall, J. Herbst, R. Müller, and H. D. Schotten: “Benchmarking OpenWiFiSync on ESP32: Towards Cost-Effective Wireless Time Synchronization”. In: 30th IEEE International Conference on Emerging Technologies and Factory Automation (ETFA), IEEE, 2025.}
}
%
\input{./organization/IEEE_authors-long.tex}%
%
%
%
%
%
%
%
\maketitle
%
%
%
%
%
\begin{abstract}%
Wireless time synchronization of mobile devices is a key enabler for numerous Industry 4.0 applications, such as coordinated and synchronized tasks or the generation of high-precision timestamps for machine learning or artificial intelligence algorithms. Traditional wireline clock synchronization protocols, however, cannot achieve the performance in wireless environments without significant modifications. To address this challenge, we make use of the Reference Broadcast Infrastructure Synchronization protocol, which leverages the broadcast nature of wireless communications and remains both non-invasive and standard-compliant. We implement and validate this protocol on a low-cost testbed using ESP32 modules and a commercial \mbox{Wi-Fi} access point. To support further research and development, we release our implementation as open-source software under the GNU General Public License Version 3 license via the OpenWifiSync project on GitHub\footnote{https://github.com/dfki-in/OpenWiFiSync/tree/esp32-rbis}. 

Our results demonstrate that synchronization accuracies within ±30 microseconds are achievable using energy-efficient and affordable hardware, making this approach suitable for a wide range of use cases.

\end{abstract}%
%
%
%
%
%
\IEEEpeerreviewmaketitle
%
%
%
%
%
%
%
%
\tikzstyle{descript} = [text = black,align=center, minimum height=1.8cm, align=center, outer sep=0pt,font = \footnotesize]
\tikzstyle{activity} =[align=center,outer sep=1pt]

\section{Introduction}%
\label{sec:Introduction}
A common and precise understanding of time is essential for both human interaction and the operation of computing devices on earth and beyond~\cite{7772527}. While humans can typically tolerate timing deviations within the range of a few seconds, devices and computer systems require significantly higher temporal accuracy. This need becomes especially pronounced in scenarios involving coordinated tasks between devices, such as robots, where timing precision directly impacts the speed and precision of a task.

To address these requirements, real-time classes have been introduced, aiming for synchronization accuracies below one microsecond~\cite{7883994}. Importantly, not only the end devices but also infrastructure components must maintain high-precision synchronization to ensure the proper functioning of time-sensitive systems. This has led to the development of various \glspl{csp}, each tailored to meet specific application dependent requirements.

However, the time synchronization of wireless systems presents unique challenges. Unlike wired systems, wireless communication is inherently more dynamic and susceptible to interference, making it difficult to directly apply existing \glspl{csp} developed for wired environments. As a result, these protocols often underperform when applied to wireless setups due to their inability to cope with the dynamic nature of such environments.


To leverage the specific characteristics of wireless systems rather than be hindered by them, the \gls{rbis} protocol has been proposed as an efficient solution for synchronizing end devices in infrastructure-based wireless systems such as \mbox{Wi-Fi}~\cite{6489696}. Previous studies have demonstrated that \gls{rbis} as well as other protocols can achieve high accuracy when supported by specialized hardware and access to hardware-level timestamping~\cite{7782431}. However, existing implementations are not publicly available and are limited to proprietary or modified hardware platforms.

To address this limitation, the OpenWiFiSync project was initiated. This project is available as \gls{oss} under \gls{gplv3} license on GitHub and aims to make accurate time synchronization accessible to the broader research community by also targeting \gls{cots} hardware. Nevertheless, the performance of \glspl{csp} on standard computing platforms is inherently limited by their general-purpose operating systems and non-deterministic processing chains~\cite{10710826}.

Unlike traditional computers, embedded devices can often achieve higher synchronization accuracy due to their deterministic hardware and software environments. In recent years, ESP32-based microcontrollers have gained popularity for \gls{iot} applications due to their compact form factor, energy efficiency, low cost, and adequate performance. This makes them well suited for human‐centric \gls{iot} as well as \gls{iiot} applications~\cite{herbst_ring_2024}.

These qualities make them promising candidates for extending the capabilities of OpenWiFiSync. In this paper, we present two key contributions:
\begin{itemize}
  \item Extension of the OpenWiFiSync project to support ESP32-based microcontrollers.
  \item Benchmarking the performance of the \gls{rbis} protocol on this platform.
\end{itemize}
Through these contributions, we aim to demonstrate that accurate wireless time synchronization can be achieved using accessible, off-the-shelf embedded hardware.

Accordingly, the paper is structured as follows: Section~\ref{sec:Related Work} outlines related work. Furthermore, the \gls{rbis} protocol is introduced in  Section~\ref{sec:rbis}. Then, the experimental setup is introduced in Section~\ref{sec:Experimental Setup} and evaluated in Section~\ref{sec:Evaluation}. Then, a outlook on the next steps of the OpenWiFiSync project is provided in Section~\ref{sec:Outlook}. 
Finally, Section~\ref{sec:Conclusion} concludes the paper.

\section{Related Work}%
\label{sec:Related Work}
Time synchronization of wireless devices has been investigated for decades, but it has gained momentum in recent years, driven by the increasing number of wireless use cases within the \gls{iiot}. To explore different \glspl{csp}, \textit{Mahmood et al.}~\cite{7782431} conducted a survey of \glspl{csp} available for \mbox{Wi-Fi}. Among the benchmarked \glspl{csp} is the so-called \gls{rbis} protocol. It was first introduced by \textit{Cena et al.}~\cite{6489696} in several configurations, with the most promising configuration evaluated in detail~\cite{7018946}. The protocol is particularly suitable as it complies with existing standards and can utilize current infrastructure components, such as \glspl{ap} for \mbox{Wi-Fi}. To emphasize the relevance of this topic, \textit{Gundall et al.} applied the \gls{rbis} protocol in a case study 
and proposed an approach to extend its applicability~\cite{9468146}. Since no implementation of the \gls{rbis} protocol was available as \gls{oss}, the OpenWiFiSync project was initiated~\cite{10710826}. It is designed to run on \gls{cots} hardware, although it remains limited in achievable performance. Therefore, this paper also investigates the use of embedded devices to overcome these limitations.

\section{Reference Broadcast Infrastructure Synchronization Protocol} %
\begin{figure}[htbp]
\centering
 \includegraphics[scale=0.9]{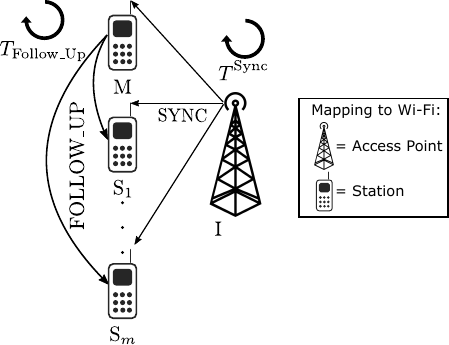}
\caption{Visualization of the RBIS protocol (refined from \cite{7018946}).}
\label{fig:rbis}
\end{figure}

\label{sec:rbis}
Wireless channel characteristics enable the concurrent reception of broadcast messages by multiple end devices. This inherent property has motivated the exploration of one-way message exchange mechanisms for \gls{csp}. Thus, this section presents the \gls{rbis} protocol, which makes use of these characteristics. Although originally proposed in the context of \mbox{Wi-Fi}, its principles are not limited to any specific wireless technology, as long as the technology is infrastructure-based. Consequently, the components illustrated in Figure~\ref{fig:rbis} are intentionally presented in a generic manner to reflect this broad applicability. 
To aid interpretation, a mapping to the \mbox{Wi-Fi} components is also provided. As shown in Figure~\ref{fig:rbis}, the infrastructure periodically emits a broadcast signal, referred to as SYNC, which is received by all devices within the communication range. In \mbox{Wi-Fi} networks, beacon frames can serve as SYNC messages. These frames include the SSID of the \gls{ap}, the transmission interval, and a timestamp representing the elapsed time since the \gls{ap} was powered on. While the default transmission interval is 102.4 ms, this value can be adjusted as needed. Upon receiving the SYNC signal, each device records a timestamp based on its local clock.

Subsequently, the master node transmits its \grqq correct" timestamps, denoted by $T_\mathrm{M}(k)$, to all slave nodes via the \mbox{FOLLOW\_UP} message. Each slave then generates corresponding reference timestamps, denoted by $T_\mathrm{S}(k)$. These timestamp pairs, $(T_\mathrm{M}(k), T_\mathrm{S}(k))$, are used to estimate the clock offset $\theta$ and clock skew $\gamma$ between the master and slave clocks, according to Equations~\ref{eq:1}--\ref{eq:2}~\cite{6817598}:
\begin{equation}
\hat{\theta}(k)=T_\mathrm{S}(k)-{T}_\mathrm{M}(k)
\label{eq:1}
\end{equation}
\begin{equation}
\hat{\gamma}(k)=\frac{\hat{\theta}(k)-\hat{\theta}(k-1)}{{T}_\mathrm{M}(k)-{T}_\mathrm{M}(k-1)}
\label{eq:2}
\end{equation}
Notably, the intervals of the SYNC and FOLLOW\_UP messages are not required to be identical.

\section{Experimental Setup}%
\label{sec:Experimental Setup}
This section describes the experimental setup including the hardware that was chosen for evaluating \gls{rbis} protocol using \mbox{Wi-Fi}. As already mentioned, a low-cost and low‐power hardware platform was targeted. Hence, ESP32‐S3 was selected (see  Table~\ref{tab:hardware}).
\begin{table}[htbp]
    \centering
    \caption{Hardware specification.}
    \begin{tabularx}{\columnwidth}{c   c   X}
    \toprule
    \textbf{Equipment} & \textbf{QTY} & \textbf{Specification}\\
    \midrule
    ESP32 & 3 & Espressif ESP32-S3 Dev-Board, Tensilica LX7 processor, 80-240 MHz  \\
    Wi-Fi Router & 1 & Linksys Broadband Model WRT54GL, IEEE 802.11n\\
      \bottomrule 
    \end{tabularx}
    \label{tab:hardware}
\end{table}
It is a versatile microcontroller that features a dual‐core Tensilica LX7 processor running at a maximum rate of 240~MHz, an onboard ultra-low‐power coprocessor, and an integrated 2.4 GHz radio. In given implementation, each ESP32‐S3 operates in \mbox{Wi-Fi} promiscuous (monitor) mode to capture raw 802.11 beacon frames broadcast by a standard \gls{ap}. The arrival times of these beacons are then used to synchronize the clock of the Slave-ESP to the Master-ESP in accordance to the \gls{rbis} protocol. As the base clock value cannot be changed, the gp-timer peripheral is used to create a synchronisable clock with an accuracy of 1\,µs, which is then still based on the ESP's hardware clock (the underlying oscillator has a frequency stability of around 10 ppm at 40\,MHz). 
\begin{figure}[htbp]
\centering
 \includegraphics[width=0.35\textwidth]{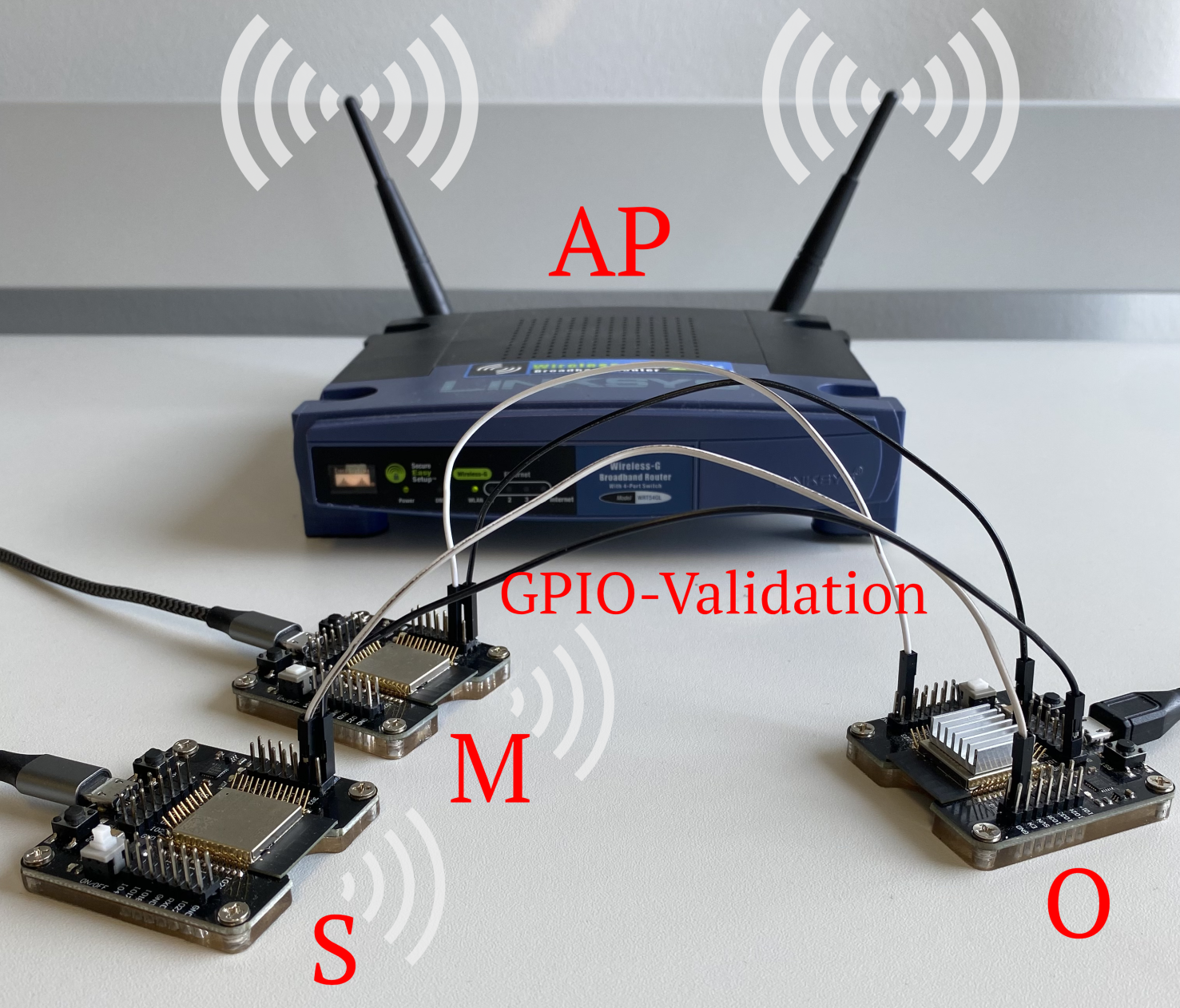}
\caption{Illustration of the experimental setup including three ESP32, a \mbox{Wi-Fi} router as well as the validation concept using GPIO pins.}
\label{fig:OverviewSetup}
\end{figure}
In Figure~\ref{fig:OverviewSetup} the experimental setup is shown, including the mentioned \gls{cots} \gls{ap}, two ESP32 microcontrollers taking on the role as master (M) and slave (S) (as in Figure~\ref{fig:rbis}) and one additional ESP32 working as an observer~(O). This last node's task is to collect and log the estimated values for $\theta$ and $\gamma$, both of which are calculated on node S and periodically sent out over \mbox{Wi-Fi}. Both M and S are connected to O via one GPIO each and additionally via GND. These GPIOs are toggled at periodic, master-relative timestamps. The time difference between the toggles by M and S can then be used to obtain measurements of $\theta$ for validation of the estimated values. While listening to the beacons in monitor mode, the ESP32‐S3 nodes associate with the same \gls{ap} in normal station mode, creating a conventional  \mbox{Wi-Fi} link for subsequent payload exchanges. By applying \gls{rbis} protocol, each node can synchronize its local clock based on beacon‐derived timestamps while still using the router’s \mbox{Wi-Fi} network for exchanging FOLLOW\_UP messages.

\section{Evaluation}%
\label{sec:Evaluation}
To benchmark the performance of the OpenWiFiSync project on ESP32-based microcontrollers, a measurement series consisting of 6,000 beacon frames was conducted, corresponding to an observation period of approximately 10 minutes. 

The evaluation focuses on several metrics: the estimated clock offset $\hat{\theta}$ and estimated clock skew $\hat{\gamma}$, both derived from the timestamps collected during protocol execution. These metrics were calculated using Equations~\ref{eq:1}\&~\ref{eq:2}. To assess the accuracy of the estimated offset, a hardware-based verification circuit was employed. As mentioned before, a GPIO pin of the master is set and detected on the slave node. Thus, the actual clock offset $\theta$ is identified and a comparison with the estimations can be done. 

\begin{figure}[htbp]
	\centering
		\subfloat[Histogram of the probability density functions obtained for the readings of $\hat{\theta}$.]{\resizebox{0.9\columnwidth}{!}{\input{figures/offsets_hist.tikz}}\label{fig:1}}
        
        \subfloat[Histogram of the probability density functions obtained for the readings of $\hat{\gamma}$.]{\resizebox{0.9\columnwidth}{!}{\input{figures/skew_hist.tikz}}\label{fig:2}}

		\subfloat[Time series obtained for the readings of $\theta$.]{\resizebox{0.9\columnwidth}{!}{\input{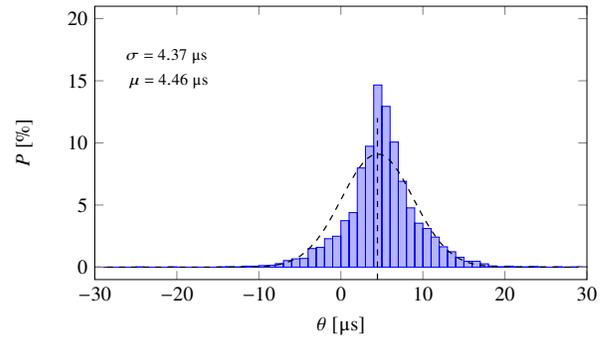}
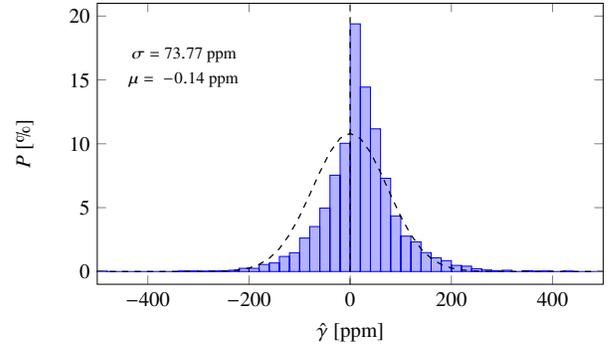
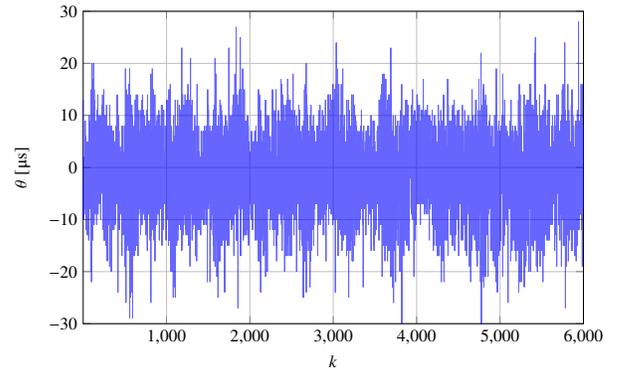
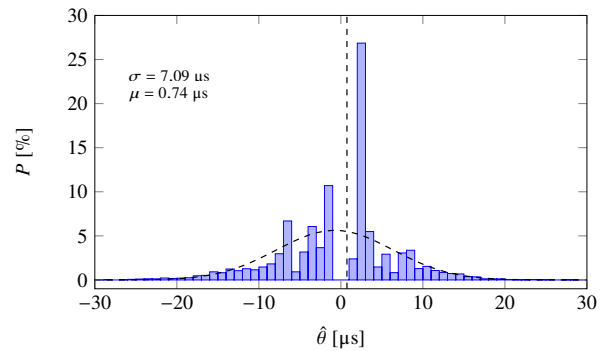}\label{fig:3}}
        
        \subfloat[Histogram of the probability density functions obtained for the readings of $\theta$.]{\resizebox{0.9\columnwidth}{!}{\input{figures/validation_hist.tikz}}\label{fig:4}}
\caption{Overview of measurement results, while Figure~\ref{fig:1},~\ref{fig:2}, and~\ref{fig:4} show histograms of the observations, Figure~\ref{fig:3} presents the readings as time series.}
\label{fig:results}
\end{figure}

The results of the measurement series is shown in Figure~\ref{fig:results}.
Here, Figure~\ref{fig:1},~\ref{fig:2}, and~\ref{fig:4} show histograms of the observations, and Figure~\ref{fig:3} presents the readings of the real offset as time series.
As illustrated in Figure~\ref{fig:1}, the estimated clock offset follows approximately a normal distribution, with most values falling within a range of ±30 microseconds. However, the distribution is slightly shifted by approximately +4 microseconds, suggesting either a consistent delay in timestamp generation on one of the ESP32 devices, or a systematic clock skew occurring within a single beacon interval. This hypothesis is supported by the results in Figure~\ref{fig:2} that shows the clock skew. The magnitude of this skew is relatively high when compared to typical values observed in \gls{cots} CPUs~\cite{10710826}, but remains acceptable considering the constraints of low-cost embedded systems.

Since the offset and skew estimations are based on observed timestamps and may not reflect the absolute timing errors, a verification setup was used to assess the actual offset. The results of this measurements are presented in Figure~\ref{fig:3} and Figure~\ref{fig:4}, where Figure~\ref{fig:3} shows a time series of synchronization offsets and Figure~\ref{fig:4} displays the corresponding histogram.

\begin{table}[htbp]
    \centering
    \caption{Measured clock precision for different standard deviations.}
    \begin{tabular}{c   c   c   c}
    \toprule
   & $\boldsymbol{\sigma}$ & $2\boldsymbol{\sigma}$  & $3\boldsymbol{\sigma}$ \\
     \midrule
  $E(\theta=\upmu)$ [$\upmu$s]   & 0.74$\pm$7.09  & 0.74$\pm$14.18 & 0.74$\pm$21.27 \\
   $P [\mathrm{\%}]$ & 68.27& 95.45& 99.73\\
      \bottomrule 
    \end{tabular}
    \label{tab:probabilities}
\end{table}
From the validation data, it is evident that the mean synchronization offset is close to zero, and the maximum observed values remain within ±25 microseconds. The standard deviation of the offset is approximately 5.35 microseconds, with over 90\% of the values falling within ±15 microseconds and 99\% within ±22 microseconds.

While other works in the field have demonstrated higher precision, the achieved results represent very good synchronization accuracy for low-cost, embedded ESP32 microcontrollers, especially considering their limited processing power. This confirms that OpenWiFiSync, in combination with the \gls{rbis} protocol, is a viable solution for accurate time synchronization on \gls{cots} embedded devices.

\section{Outlook on OpenWiFiSync Project}%
\label{sec:Outlook}
The ongoing development of the OpenWiFiSync project includes several key directions aimed at enhancing both the performance and applicability. Here, a first improvement is the application of rate correction techniques. Therefore, moving average filters, Kalman filters, and advanced control-theoretic methods such as Linear Quadratic Gaussian (LQG) Linear Quadratic Regulator (LQR) control can be used. 

Another important objective is the expansion of supported hardware platforms. The initial implementation targeted a high-performance mini-PC as the receiver device. More recently, the protocol has been successfully ported to resource-constrained platforms such as the ESP32, demonstrating its feasibility even under limited computational resources. Future efforts aim to bridge the performance gap by supporting embedded systems with moderate capabilities and \gls{cots} operating systems, such as the Raspberry Pi or LattePanda. 


The applicability of the protocol to infrastructure-less networks is also under active consideration. In scenarios such as Wireless Body Area Networks (WBANs), where centralized infrastructure is unavailable, relative synchronization becomes essential~\cite{herbst_ring_2024}. For these use cases, the Reference Broadcast Synchronization (RBS) approach can offer a effective solution.


\section{Conclusion}%
\label{sec:Conclusion}
Wireless time synchronization is a key enabler for a wide range of future industrial applications, particularly in the context of the \gls{iiot}. Low-cost and energy-efficient devices, such as ESP32 microcontrollers, are expected to play a significant role in these scenarios. In this paper, we investigated the applicability of the \gls{rbis} protocol on such hardware platforms. Our work is part of the OpenWifiSync project, and all results and implementations are made available as \gls{oss} under the \gls{gplv3} license on GitHub.

Experimental results show that an accuracy of ±30 microseconds can be achieved using offset correction, demonstrating the feasibility of precise wireless synchronization even with constrained hardware. While this level of accuracy is already sufficient for many \gls{iiot} applications, further improvements are possible. 




\printbibliography%

%
%
\end{document}

%% file: organization/preamble.tex
\PassOptionsToPackage{usenames,dvipsnames,svgnames,x11names,table,prologue}{xcolor}%
\PassOptionsToPackage{hyphens}{url}%
\PassOptionsToPackage{nomessages}{fp}%
\ifCLASSINFOpdf
  \usepackage[pdftex]{graphicx}
\else
  \usepackage[dvips]{graphicx}
\fi
\ifCLASSOPTIONcompsoc
   \usepackage[caption=false,font=normalsize,labelfont=sf,textfont=sf]{subfig}
\else
   \usepackage[caption=false,font=footnotesize]{subfig}
\fi

\usepackage{stfloats}
\usepackage[english]{babel}%
\usepackage[utf8]{inputenc}%
\usepackage[babel,style=english]{csquotes}%
\usepackage{hyphsubst}%
\usepackage[%
	activate={true,%
	nocompatibility},%
	final,%
	tracking=true,%
	kerning=true,%
	spacing=true,%
	factor=1100,%
	stretch=10,%
	shrink=10%
]{microtype}%
\usepackage{setspace}%
\usepackage{xcolor}%
\definecolor{todonotecol}{RGB}{250,0,0}%
\usepackage{xparse}
\usepackage[%
	colorlinks=false,%
	urlcolor=black,%
	linkcolor=black,%
	citecolor=black,%
	filecolor=black,%
	breaklinks,%
	]{hyperref}%
\usepackage{url}%
\usepackage[%
	acronym,%
	nopostdot,%
	seeautonumberlist,%
	shortcuts,%
	section=chapter,%
	toc,%
]{glossaries}%
\loadglsentries{./supply/glossaries.tex}\glsdisablehyper
\usepackage[%
	backend=biber,%
	style=ieee,%
	isbn=false,%
	hyperref=true,%
	maxbibnames=99,%
	sorting=none,%
	natbib=true,%
	language=english,%
	defernumbers=true,%
	]{biblatex}%
\DeclareFieldFormat{sentencecase}{\csname bbx@colon@search\endcsname#1}

\usepackage{nameref}
\usepackage{soul}
\usepackage{flushend}
\usepackage{textcomp}
\usepackage{calc}
\usepackage{xkeyval}
\usepackage{multirow}
\usepackage{tabulary}
\usepackage{makecell}
\usepackage{pgfplots}
\usepackage{tikz}
\usepackage{textcomp} 
\usepackage{verbatim}

%% file: supply/glossaries.tex
\glsaddkey*{url}
{}
{\glsentryurl}
{\Glsentryurl}
{\glsurl}
{\Glsurl}
{\GLSurl}
\glsaddkey*{longpltwo}
{}
{\glsentrylongpltwo}
{\Glsentrylongpltwo}
{\glslongpltwo}
{\Glslongpltwo}
{\GLSlongpltwo}
%
%
%
%
%
\newglossaryentry{}{%
	name={},%
	plural={},%
	description={},%
}%
%
%
%
%
%
%
\newacronym{it}{IT}{information technology}
\newacronym{ot}{OT}{operational technology}
\newacronym{opcua}{OPC UA}{Open Platform Communications Unified Architecture}
\newacronym{tsn}{TSN}{Time-Sensitive Networking}
\newacronym{scada}{SCADA}{supervisory control and data acquisition}
\newacronym{vpc}{VPC}{Virtual Process Controller}
\newacronym{plc}{PLC}{Programmable Logic Controller}
\newacronym{udp}{UDP}{User Datagram Protocol}
\newacronym{pubsub}{PubSub}{Publish Subscribe}
\newacronym{vm}{VM}{virtual machine}
\newacronym{pc}{PC}{personal computer}
\newacronym{ptp}{PTP}{Precision Time Protocol}
\newacronym{gptp}{gPTP}{generalized Precision Time Protocol}
\newacronym{ntp}{NTP}{Network Time Protocol}
\newacronym{splc}{Soft-PLC}{Software-PLC}
\newacronym{rte}{RTE}{Real-Time Ethernet}
\newacronym{osi}{OSI}{Open Systems Interconnection}
\newacronym{mac}{MAC}{Media Access Control}
\newacronym{e2e}{E2E}{end-to-end}
\newacronym{tcp}{TCP}{Transmission Control Protocol}
\newacronym{hmi}{HMI}{Human-Machine Interface}
\newacronym{kpi}{KPI}{Key Performance Indicator}
\newacronym{mes}{MES}{Manufacturing Execution System}
\newacronym{rami4.0}{RAMI 4.0}{Reference Architectural Model Industrie 4.0}
\newacronym{tacnet4.0}{TACNET 4.0}{Tactile Internet 4.0}
\newacronym{3gpp}{3GPP}{3rd Generation Partnership Project}
\newacronym{5gppp}{5GPPP}{5G Infrastructure Public Private Partnership}
\newacronym{itu}{ITU}{International Telecommunication Union}
\newacronym{ngnm}{NGNM}{Next Generation Mobile Networks}
\newacronym{agv}{AGV}{automated guided vehicle}
\newacronym{v2v}{V2V}{vehicle-to-vehicle}
\newacronym{v2x}{V2X}{vehicle-to-infrastructure}
\newacronym{d2x}{D2X}{device-to-x}
\newacronym{qos}{QoS}{quality of service}
\newacronym{noa}{NOA}{NAMUR Open Architecture}
\newacronym{dcs}{DCS}{distributed control system}
\newacronym{m2m}{M2M}{machine-to-machine}
\newacronym{sdn}{SDN}{software-defined networking}
\newacronym{erp}{ERP}{enterprise resource planning}
\newacronym{ip}{IP}{Internet Protocol}
\newacronym{lte}{LTE}{Long Term Evolution}
\newacronym{5g}{5G}{5th generation wireless communication systems}
\newacronym{harq}{HARQ}{Hybrid Automatic Repeat Request}
\newacronym{rat}{RAT}{radio access technology}
\newacronym{mc}{MC}{Multi-Connectivity}
\newacronym{fbmc}{FBMC}{Filter Bank Multicarrier}
\newacronym{gfdm}{GFDM}{Generalized Frequency Division Multiplexing}
\newacronym{sla}{SLA}{service-level agreement}
\newacronym{5qi}{5QI}{5G QoS Indicator}
\newacronym{bmbf}{BMBF}{German Federal Ministry of Education and Research}
\newacronym{nrt}{NRT}{non real-time}
\newacronym{irt}{IRT}{isochronous real-time}
\newacronym{rt}{RT}{real-time}
\newacronym{ar}{AR}{augmented reality}
\newacronym{wlan}{WLAN}{Wireless Local Area Network}
\newacronym{cpu}{CPU}{central processing unit}
\newacronym{ram}{RAM}{random-access memory}
\newacronym{ue}{UE}{user equipment}
\newacronym{ran}{RAN}{radio access network}
\newacronym{bs}{BS}{base station}
\newacronym{cn}{CN}{core network}
\newacronym{epc}{EPC}{evolved packet core}
\newacronym{mec}{MEC}{mobile edge cloud}
\newacronym{rtt}{RTT}{Round Trip Time}
\newacronym{vpn}{VPN}{virtual private network}
\newacronym{vpf}{VPF}{Virtual Process Control Function}
\newacronym{vdi}{VDI}{Association of German Engineers}
\newacronym{ipc}{IPC}{industrial pc}
\newacronym{rtos}{RTOS}{real-time operating system}
\newacronym{mano}{VPC M\&O}{VPC Manager \& Orchestrator}
\newacronym{i/o}{I/O}{Input / Output}
\newacronym{fbd}{FBD}{Function Block Diagram}
\newacronym{ldd}{LDD}{Ladder Logic Diagram}
\newacronym{sfc}{SFC}{Sequential Function Chart}
\newacronym{il}{IL}{Instruction List}
\newacronym{st}{ST}{Structured Text}
\newacronym{os}{OS}{Operating System}
\newacronym{mqtt}{MQTT}{message queuing telemetry transport }
\newacronym{amqp}{AMQP}{advanced message queuing protocol}
\newacronym{api}{API}{application programming interface}
\newacronym{embb}{eMBB}{enhanced mobile broadband}
\newacronym{iot}{IoT}{Internet of Things}
\newacronym{urllc}{URLLC}{ultra-reliable low-latency communication}
\newacronym{mmtc}{mMTC}{massive machine-type communications}
\newacronym{iiot}{IIoT}{Industrial IoT}
\newacronym{iiot2}{IIoT}{Industrial Internet of Things}
\newacronym{npn}{NPN}{non-public network}
\newacronym{prb}{PRB}{Physical Ressource Block}
\newacronym{gnb}{gNB}{5G base station}
\newacronym{5gs}{5GS}{5G system}
\newacronym{5gc}{5GC}{5G core network}
\newacronym{oai}{OAI}{OpenAirInterface}
\newacronym{enb}{eNB}{4G base station}
\newacronym{sdr}{SDR}{Software Defined Radio}
\newacronym{usrp}{USRP}{Universal Software Radio Peripheral}
\newacronym{mme}{MME}{Mobility Management Entity}
\newacronym{hss}{HSS}{Home Subscriber Server}
\newacronym{pdn}{PDN}{Packet Data Network}
\newacronym{sgw}{SGW}{Serving Gateway}
\newacronym{gtp}{GTP}{GPRS Tunneling Protocol}
\newacronym{mimo}{MIMO}{Multiple Input Multiple Output}
\newacronym{fpga}{FPGA}{Field Programmable Gate Array}
\newacronym{pss}{PSS}{primary synchronization signal}
\newacronym{sss}{SSS}{secondary synchronization signal}
\newacronym{sfn}{SFN}{system frame number}
\newacronym{mib}{MIB}{master information block}
\newacronym{pbch}{PBCH}{Physical Broadcast Channel}
\newacronym{ism}{ISM}{Industrial, Scientific and Medical}
\newacronym{e-utra}{E-UTRA}{evolved UMTS Terrestrial Radio Access}
\newacronym{bldc}{BLDC}{brushless DC}
\newacronym{dl}{DL}{Downlink}
\newacronym{ds-tt}{DS-TT}{Device-Side TSN Translator}
\newacronym{nw-tt}{NW-TT}{Network-Side TSN Translator}
\newacronym{upf}{UPF}{User Plane Function}
\newacronym{tsi}{TSi}{ingress timestamping}
\newacronym{tse}{TSe}{egress timestamping}
\newacronym{gm}{GM}{grandmaster}
\newacronym{sn}{SN}{sequence number}
\newacronym{rbis}{RBIS}{Reference Broadcast Infrastructure Synchronization}
\newacronym{cococo}{CoCoCo}{communication-control codesign}
\newacronym{3c}{3C}{joint design of communications, control, and computing}
\newacronym{cots}{COTS}{Commercial Off-The-Shelf}
\newacronym{ie}{IE}{industrial Ethernet}
\newacronym{ap}{AP}{access point}
\newacronym{ofdm}{OFDM}{Orthogonal Frequency-Division Multiplexing}
\newacronym{csma}{CSMA-CA}{carrier sense multiple access with collision avoidance}
\newacronym{dcf}{DCF}{distributed coordination function}
\newacronym{pcf}{PCF}{point coordination function}
\newacronym{hcf}{HCF}{hybrid coordination function}
\newacronym{edca}{EDCA}{enhanced distributed channel access}
\newacronym{hcca}{HCCA}{\gls{hcf} controlled channel access}
\newacronym{tdma}{TDMA}{time division multiple access}
\newacronym{ofdma}{OFDMA}{Orthogonal Frequency-Division Multiple Access}
\newacronym{mumimo}{MU-MIMO}{multi-user \gls{mimo}}
\newacronym{bi}{BI}{beacon interval}
\newacronym{ssid}{SSID}{service set identifier}
\newacronym{csp}{CSP}{Clock Synchronization Protocol}
\newacronym{oss}{OSS}{Open Source Software}
\newacronym{gplv3}{GNU GPLv3}{GNU General Public License Version 3}

%% file: organization/makros.tex
\newcommand{\mytilde}{{\raise.17ex\hbox{$\scriptstyle\mathtt{\sim}$}}}

%% file: organization/settings.tex
\newlength\textheighttemp%
\newlength\textwidthtemp%
\newlength\textheightstd%
\newlength\textwidthstd%
\newlength\textheightold%
\newlength\textwidthold%
\newlength\tempheight%
\newlength\tempwidth%
\SetProtrusion{encoding={*},family={bch},series={*},size={6,7}}
              {1={ ,750},2={ ,500},3={ ,500},4={ ,500},5={ ,500},
               6={ ,500},7={ ,600},8={ ,500},9={ ,500},0={ ,500}}
\SetExtraKerning[unit=space]
    {encoding={*}, family={bch}, series={*}, size={footnotesize,small,normalsize}}
    {\textendash={400,400}, 
     "28={ ,150}, 
     "29={150, }, 
     \textquotedblleft={ ,150}, 
     \textquotedblright={150, }} 
\SetTracking{encoding={*}, shape=sc}{40}

%% file: organization/IEEE_authors-long.tex
\author{%
\IEEEauthorblockN{%
    Michael Gundall\IEEEauthorrefmark{1}, %
    Jan Herbst\IEEEauthorrefmark{1},
    Robin Müller\IEEEauthorrefmark{1},
    and Hans D. Schotten\IEEEauthorrefmark{1}\IEEEauthorrefmark{3} %
    \\%
}%
\IEEEauthorblockA{%
    \IEEEauthorrefmark{1}German Research Center for Artificial Intelligence (DFKI), Kaiserslautern, Germany \\%
    \IEEEauthorrefmark{3}Department of Electrical and Computer Engineering, RPTU University Kaiserslautern-Landau, Kaiserslautern, Germany %
	\\%
    Email: %
        \{michael.gundall, jan.herbst, robin.mueller, hans\_dieter.schotten%
        \}@dfki.de
        \\%
}%
}%

%% file: figures/offsets_hist.tikz
\definecolor{mycolor1}{rgb}{0.00000,0.3,0.6}
\definecolor{mycolor3}{rgb}{0.01,0.79,0.395}

\begin{tikzpicture}
\begin{axis}[ 
width=4in,
height=2.5in,
xmin=-30, xmax=30,
ymin=-1, ymax=21,
area style,
ytick={0,5,10,15,20},
ylabel={$P$ [$\mathrm{\%}$]},
xlabel={$\theta$ [$\upmu$s]},
    ]
\addplot+[ybar interval,mark=no] plot coordinates { 
(	-30	,	0	)
(	-29	,	0	)
(	-28	,	0	)
(	-27	,	0	)
(	-26	,	0	)
(	-25	,	0.02	)
(	-24	,	0	)
(	-23	,	0	)
(	-22	,	0	)
(	-21	,	0.02	)
(	-20	,	0	)
(	-19	,	0	)
(	-18	,	0	)
(	-17	,	0	)
(	-16	,	0	)
(	-15	,	0.02	)
(	-14	,	0.03	)
(	-13	,	0.02	)
(	-12	,	0.08	)
(	-11	,	0.08	)
(	-10	,	0.13	)
(	-9	,	0.15	)
(	-8	,	0.28	)
(	-7	,	0.53	)
(	-6	,	0.67	)
(	-5	,	0.7	)
(	-4	,	1.5	)
(	-3	,	1.59	)
(	-2	,	2.29	)
(	-1	,	2.49	)
(	0	,	3.74	)
(	1	,	4.39	)
(	2	,	7.98	)
(	3	,	9.73	)
(	4	,	14.66	)
(	5	,	12.94	)
(	6	,	10.07	)
(	7	,	6.91	)
(	8	,	4.77	)
(	9	,	3.54	)
(	10	,	3.11	)
(	11	,	2.42	)
(	12	,	1.64	)
(	13	,	1.24	)
(	14	,	0.8	)
(	15	,	0.47	)
(	16	,	0.47	)
(	17	,	0.27	)
(	18	,	0.1	)
(	19	,	0.03	)
(	20	,	0.08	)
(	21	,	0.03	)
(	22	,	0.02	)
(	23	,	0.05	)
(	24	,	0.02	)
(	25	,	0	)
(	26	,	0.03	)
(	27	,	0	)
(	28	,	0.02	)
(	29	,	0.02	)
(	30	,	0	)
};

\addplot [color=black, style={semithick}, dashed, forget plot]
  table[row sep=crcr]{%
-30		2.67E-13	\\
-29		1.59E-12	\\
-28		8.96E-12	\\
-27		4.80E-11	\\
-26		2.44E-10	\\
-25		1.17E-09	\\
-24		5.36E-09	\\
-23		2.33E-08	\\
-22		9.58E-08	\\
-21		3.74E-07	\\
-20		1.39E-06	\\
-19		4.88E-06	\\
-18		1.63E-05	\\
-17		5.15E-05	\\
-16		0.000154813	\\
-15		0.000441341	\\
-14		0.001193847	\\
-13		0.00306429	\\
-12		0.007463081	\\
-11		0.017247	\\
-10		0.037819522	\\
-9		0.078691102	\\
-8		0.155361111	\\
-7		0.291049	\\
-6		0.517364955	\\
-5		0.872639779	\\
-4		1.396625954	\\
-3		2.120959197	\\
-2		3.05626932	\\
-1		4.178862033	\\
0		5.421650864	\\
1		6.674398979	\\
2		7.796503609	\\
3		8.641610894	\\
4		9.088592539	\\
5		9.069966075	\\
6		8.588588492	\\
7		7.716938335	\\
8		6.57923453	\\
9		5.322464937	\\
10		4.08561419	\\
11		2.975836048	\\
12		2.056684857	\\
13		1.348756664	\\
14		0.839279392	\\
15		0.495549017	\\
16		0.277634723	\\
17		0.147593777	\\
18		0.074450808	\\
19		0.035635092	\\
20		0.016184283	\\
21		0.006974549	\\
22		2.85E-03	\\
23		0.00110658	\\
24		0.000407405	\\
25		0.000142324	\\
26		4.72E-05	\\
27		1.48E-05	\\
28		4.43E-06	\\
29		1.25E-06	\\
30		3.37E-07	\\
};
\node[at={(87,180)},fill= white, text=black] {\footnotesize{$\sigma=4.37~\upmu \mathrm{s}$}};
\node[at={(90,160)},fill= white, text=black] {\footnotesize{$\mu=4.46~\upmu \mathrm{s}$}};
 \addplot [color=black,dashed, style={semithick}, forget plot]
  table[row sep=crcr]{%
 4.46   -1 \\
  4.46  12\\
 };
\end{axis}
\end{tikzpicture}

%% file: figures/skew_hist.tikz
\definecolor{mycolor1}{rgb}{0.00000,0.3,0.6}
\definecolor{mycolor3}{rgb}{0.01,0.79,0.395}

\begin{tikzpicture}
\begin{axis}[ 
width=4in,
height=2.5in,
xmin=-500, xmax=500,
area style,
ymin=-1, ymax=21,
ytick={0,5,10,15,20},
ylabel={$P$ [$\mathrm{\%}$]},
xlabel={$\hat{\gamma}$ [ppm]}
    ]
\addplot+[ybar interval,mark=no] plot coordinates { 
(	-500	,	0.02	)
(	-480	,	0.00E+00	)
(	-460	,	0	)
(	-440	,	0	)
(	-420	,	0	)
(	-400	,	0	)
(	-380	,	0	)
(	-360	,	0	)
(	-340	,	0.03	)
(	-320	,	0.03	)
(	-300	,	0.02	)
(	-280	,	0.03	)
(	-260	,	0.05	)
(	-240	,	0.1	)
(	-220	,	0.25	)
(	-200	,	0.25	)
(	-180	,	0.55	)
(	-160	,	0.67	)
(	-140	,	1.18	)
(	-120	,	1.47	)
(	-100	,	2.62	)
(	-80	,	3.52	)
(	-60	,	4.97	)
(	-40	,	7.54	)
(	-20	,	10.04	)
(	0	,	19.39	)
(	20	,	14.44	)
(	40	,	11.17	)
(	60	,	7.3	)
(	80	,	4.35	)
(	100	,	2.77	)
(	120	,	2.32	)
(	140	,	1.48	)
(	160	,	1.07	)
(	180	,	0.85	)
(	200	,	0.48	)
(	220	,	0.42	)
(	240	,	0.22	)
(	260	,	0.12	)
(	280	,	0.05	)
(	300	,	0.1	)
(	320	,	0	)
(	340	,	0.07	)
(	360	,	0.02	)
(	380	,	0.03	)
(	400	,	0	)
(	420	,	0.03	)
(	440	,	0	)
(	460	,	0	)
(	480	,	0	)
(	500	,	0	)
};

\addplot [color=black, style={semithick}, dashed, forget plot]
  table[row sep=crcr]{%
-500		1.16E-09	\\
-480		7.01E-09	\\
-460		3.94E-08	\\
-440		2.06E-07	\\
-420		9.99E-07	\\
-400		4.51E-06	\\
-380		1.89E-05	\\
-360		7.35E-05	\\
-340		0.000266069	\\
-320		0.00089435	\\
-300		0.00279317	\\
-280		0.008105188	\\
-260		0.021852693	\\
-240		0.054742291	\\
-220		0.127413975	\\
-200		0.275541657	\\
-180		0.553647793	\\
-160		1.033608289	\\
-140		1.792893523	\\
-120		2.889542705	\\
-100		4.326929713	\\
-80		6.020141065	\\
-60		7.782329226	\\
-40		9.347352808	\\
-20		10.43142824	\\
0		10.81620772	\\
20		10.42035146	\\
40		9.327512069	\\
60		7.757564161	\\
80		5.994611424	\\
100		4.304005344	\\
120		2.871181659	\\
140		1.779609209	\\
160		1.024860422	\\
180		0.548379111	\\
200		0.272629715	\\
220		0.125933589	\\
240		0.054048802	\\
260		0.021552948	\\
280		0.007985523	\\
300		0.002749009	\\
320		0.000879276	\\
340		0.000261306	\\
360		7.22E-05	\\
380		1.85E-05	\\
400		4.41E-06	\\
420		9.77E-07	\\
440		2.01E-07	\\
460		3.85E-08	\\
480		6.83E-09	\\
500		1.13E-09	\\
};~
\node[at={(150,18)},fill= white, text=black] {\footnotesize{~~~~~$\sigma=73.77$ ppm}};
\node[at={(170,16)},fill= white, text=black] {\footnotesize{$\mu=~-0.14$ ppm}};
 \addplot [color=black,dashed, style={semithick}, forget plot]
  table[row sep=crcr]{%
  -0.14   -1 \\
  -0.14    120\\
 };
\end{axis}
\end{tikzpicture}

%% file: figures/validation_hist.tikz
\definecolor{mycolor1}{rgb}{0.00000,0.3,0.6}
\definecolor{mycolor3}{rgb}{0.01,0.79,0.395}

\begin{tikzpicture}
\begin{axis}[ 
width=4in,
height=2.5in,
ymin=-1, ymax=30,
xmin=-30, xmax=30,
area style,
ytick={0,5,10,15,20,25,30},
ylabel={$P$ [$\mathrm{\%}$]},
xlabel={$\hat{\theta}$ [$\upmu$s]}
    ]
\addplot+[ybar interval,mark=no] plot coordinates { 
(	-30	,	0.03	)
(	-29	,	3.00E-02	)
(	-28	,	0	)
(	-27	,	0.03	)
(	-26	,	0.07	)
(	-25	,	0.1	)
(	-24	,	0.13	)
(	-23	,	0.12	)
(	-22	,	0.22	)
(	-21	,	0.17	)
(	-20	,	0.2	)
(	-19	,	0.25	)
(	-18	,	0.47	)
(	-17	,	0.52	)
(	-16	,	0.92	)
(	-15	,	0.82	)
(	-14	,	1.24	)
(	-13	,	1.05	)
(	-12	,	1.27	)
(	-11	,	1.15	)
(	-10	,	1.47	)
(	-9	,	1.82	)
(	-8	,	2.97	)
(	-7	,	6.69	)
(	-6	,	0.92	)
(	-5	,	3.17	)
(	-4	,	6.06	)
(	-3	,	3.66	)
(	-2	,	10.7	)
(	-1	,	0	)
(	0	,	0	)
(	1	,	2.39	)
(	2	,	26.86	)
(	3	,	5.49	)
(	4	,	1.47	)
(	5	,	2.92	)
(	6	,	0.83	)
(	7	,	3.06	)
(	8	,	3.37	)
(	9	,	1.29	)
(	10	,	1.54	)
(	11	,	1.07	)
(	12	,	0.85	)
(	13	,	0.73	)
(	14	,	0.68	)
(	15	,	0.37	)
(	16	,	0.35	)
(	17	,	0.13	)
(	18	,	0.12	)
(	19	,	0.12	)
(	20	,	0.05	)
(	21	,	0.03	)
(	22	,	0.03	)
(	23	,	0.05	)
(	24	,	0.03	)
(	25	,	0.03	)
(	26	,	0	)
(	27	,	0.02	)
(	28	,	0.02	)
(	29	,	0	)
(	30	,	0	)
};

\addplot [color=black, style={semithick}, dashed,forget plot]
  table[row sep=crcr]{%
-30		0.001137137	\\
-29		0.002013922	\\
-28		0.003496552	\\
-27		0.00595121	\\
-26		0.009929749	\\
-25		0.016241985	\\
-24		0.026044003	\\
-23		0.040939651	\\
-22		0.063088228	\\
-21		0.095306015	\\
-20		0.14114324	\\
-19		0.204912119	\\
-18		0.291637249	\\
-17		0.406898539	\\
-16		0.556540863	\\
-15		0.746235271	\\
-14		0.980894449	\\
-13		1.263969429	\\
-12		1.596682818	\\
-11		1.977281589	\\
-10		2.400414172	\\
-9		2.856746017	\\
-8		3.332919833	\\
-7		3.811938574	\\
-6		4.274001878	\\
-5		4.697765264	\\
-4		5.061924984	\\
-3		5.346971707	\\
-2		5.536914997	\\
-1		5.620767421	\\
0		5.593596862	\\
1		5.457006858	\\
2		5.218979909	\\
3		4.893105045	\\
4		4.497293616	\\
5		4.05215206	\\
6		3.57921688	\\
7		3.099260523	\\
8		2.630849126	\\
9		2.189281413	\\
10		1.785973563	\\
11		1.428289524	\\
12		1.119760812	\\
13		0.860601437	\\
14		0.648405452	\\
15		0.47891566	\\
16		0.346768155	\\
17		0.246142816	\\
18		0.171278524	\\
19		0.116838628	\\
20		0.078133589	\\
21		0.051222042	\\
22		0.032918784	\\
23		0.020739509	\\
24		0.012809169	\\
25		0.007755525	\\
26		0.0046033	\\
27		0.002678521	\\
28		0.001527878	\\
29		0.000854378	\\
30		0.00046836	\\
};
\node[at={(90,24)},fill= white, text=black] {\footnotesize{$\sigma=7.09~\upmu \mathrm{s}$}};
\node[at={(90,22)},fill= white, text=black] {\footnotesize{$\mu=0.74~\upmu \mathrm{s} $}};
 \addplot [color=black,dashed, style={semithick}, forget plot]
  table[row sep=crcr]{%
  0.74   -1 \\
  0.74    102\\
 };
\end{axis}
\end{tikzpicture}